\documentclass[10pt,english,oneside,twocolumn,a4paper]{article}% change 'a4paper' to 'letter' to use imperial paper format.
\usepackage{graphicx}
\usepackage[utf8]{inputenc}
\usepackage{color}
\usepackage{mathptmx}
\usepackage{tabularx}
\usepackage{ragged2e}
\usepackage[singlelinecheck=false]{caption}
\usepackage[T1]{fontenc}
\usepackage[numbers]{natbib}
\usepackage{amsmath}
\usepackage{listings}
 \usepackage{xcolor}
 \definecolor{mygreen}{rgb}{0,0.6,0}
\definecolor{mygray}{rgb}{0.5,0.5,0.5}
\definecolor{mymauve}{rgb}{0.58,0,0.82}
\usepackage{caption,subcaption}
\usepackage{enumitem}
\usepackage[hyphens]{url}
\usepackage{hyperref}
\usepackage{babel}
\RequirePackage[blocks]{authblk}
\makeatletter
\let\ps@plain\ps@empty
\def\@xivpt{14pt}
\def\@sect#1#2#3#4#5#6[#7]#8{%
  \ifnum #2<2
    \null\par\vskip-15pt
  \fi
  \ifnum #2>\c@secnumdepth 
    \let\@svsec\@empty
  \else
    \refstepcounter{#1}%
    \protected@edef\@svsec{%
      \ifnum #2<4
        \hb@xt@10mm{\csname the#1\endcsname}\relax
      \else
        \hb@xt@12mm{\csname the#1\endcsname}\relax
      \fi}%
  \fi
  \@tempskipa #5\relax
  \ifdim \@tempskipa>\z@
    \begingroup
      #6{%
        \@hangfrom{\hskip #3\relax\@svsec}%
          \interlinepenalty \@M #8\@@par}%
    \endgroup
    \csname #1mark\endcsname{#7}%
    \addcontentsline{toc}{#1}{%
      \ifnum #2>\c@secnumdepth \else  
        \protect\numberline{\csname the#1\endcsname}%
      \fi 
      #7}%
  \else
    \def\@svsechd{%
      #6{\hskip #3\relax
      \@svsec #8}%
      \csname #1mark\endcsname{#7}%
      \addcontentsline{toc}{#1}{%
        \ifnum #2>\c@secnumdepth \else
          \protect\numberline{\csname the#1\endcsname}%
        \fi
        #7}}%
  \fi
  \@xsect{#5}}
\renewcommand\LARGE{\@setfontsize\LARGE{16}{20}}
\def\abstract#1{\def\@abstract{#1}}
\def\abstractEn#1{\def\@abstractEn{#1}}
\def\titleEn#1{\def\@titleEn{#1}}
\def\@maketitle{%
  \newpage
  \null
  \let \footnote \thanks
    {\LARGE\bfseries\RaggedRight \@title \par}%
    \vskip 1\baselineskip%
    {\normalsize
      \@author\par}%
    \vskip 2\baselineskip%
    \vskip \baselineskip%
    {\section*{Abstract}
      \@abstract}%
  \par
  \vskip 3\baselineskip}

\renewcommand\section{\@startsection {section}{1}{\z@}%
                                   {-3.5ex \@plus -1ex \@minus -.2ex}%
                                   {\baselineskip}%
                                   {\normalfont\Large\bfseries\RaggedRight}}
\renewcommand\subsection{\@startsection{subsection}{2}{\z@}%
                                     {\baselineskip}%
                                     {1ex}%
                                     {\normalfont\large\bfseries\RaggedRight}}
\renewcommand\subsubsection{\@startsection{subsubsection}{3}{\z@}%
                                     {1\baselineskip}%
                                     {3bp}%
                                     {\normalfont\normalsize\bfseries\RaggedRight}}
\renewcommand\paragraph{\@startsection{paragraph}{4}{\z@}%
                                    {1\baselineskip\@plus1ex \@minus.2ex}%
                                    {3bp}%
                                    {\normalfont\normalsize\RaggedRight}}
\renewcommand\subparagraph{\@startsection{subparagraph}{5}{\parindent}%
                                       {3.25ex \@plus1ex \@minus .2ex}%
                                       {-1em}%
                                      {\normalfont\normalsize\bfseries\RaggedRight}}
\parindent\p@
\parskip\z@
\DeclareCaptionLabelSeparator{enskip}{\enskip}
\makeatother

\title{Run-time Performance Monitoring of Heterogenous Hw/Sw Platforms Using PAPI}

\author[a]{Tiziana Fanni}
\author[b]{Daniel Madro\~nal}
\author[c,d]{Claudio Rubattu}
\author[a]{Carlo Sau}
\author[c]{Francesca Palumbo}
\author[b]{Eduardo Ju\'arez}
\author[d]{Maxime Pelcat}
\author[b]{C\'esar Sanz}
\author[a]{Luigi Raffo}
\affil[a]{Department of Electric and Electronic Engineering, Universit\`a degli Studi di Cagliari}
\affil[b]{Research Center on Software Technologies and Multimedia Systems, Univesidad Polit\'ecnica de Madrid}
\affil[c]{Department of Chemistry and Pharmacy, Universit\`a degli Studi di Sassari}
\affil[d]{Univ Rennes, INSA Rennes, IETR UMR CNRS 6164}

\abstract{In the era of Cyber Physical Systems, designers need to offer support for run-time adaptivity considering different constraints, including the internal status of the system. This work presents a run-time monitoring approach, based on the Performance Application Programming Interface, that offers a unified interface to transparently access both the standard Performance Monitoring Counters (PMCs) in the CPUs and the custom ones integrated into hardware accelerators. Automatic tools offer to Sw programmers the support to design and implement Coarse-Grain Virtual Reconfigurable Circuits, instrumented with custom PMCs. This approach has been validated on a heterogeneous application for image/video processing with an overhead of 6\% of the execution time.}

\begin{document}

\maketitle

\section{Context and Objectives}
\label{s:intro}

Cyber-Physical Systems (CPS) are complex systems, composed of different components characterized by a strong interaction with environment and users. In particular, they need to adapt their behaviour according to the environment, any user requests and also their internal status~\cite{kim_2012}. The H2020 CERBERO European Project~\cite{masin_2017,palumbo_2019} is developing a continuous design environment for CPS, relying on a set of tools developed by project partners. Effective  support for run-time adaptation in heterogeneous systems, taking into account a plethora of different internal and external triggers, is among  the CERBERO expected outcomes, and a fundamental step is monitoring the hardware (Hw) and software (Sw) elements of the heterogeneous system~\cite{Palumbo_2019x2}. 

This paper focuses on one fundamental step necessary to design self-adaptive systems: the monitoring of heterogeneous architectures, where processing cores are connected to custom hardware accelerators that can be reconfigured at run-time. One of the Hw reconfigurable infrastructures supported in CERBERO is the Coarse-Grain Virtual Reconfigurable Circuits (CG-VRCs)~\cite{Wijtvliet_2016}. CG-VRCs offer fast and low power reconfiguration, with a good trade-off between performance and flexibility, being suitable for providing run-time Hw adaptation. In these kinds of systems, all the resources belonging to all the configurations are instantiated in the substrate and different configurations are enabled by multiplexing resources in time~\cite{hartenstein2001:1}, they can be implemented on both Field Programmable Gate Array (FPGA) or Application Specific Integrated Circuit (ASIC) systems. These kinds of accelerators are suited to support:  

\begin{enumerate}[leftmargin=*]
    \item \textit{Functional oriented adaptivity}: the application is able to execute different functionalities over the same substrate (e.g., algorithm changes)~\cite{Palumbo_2017}.
    \item \textit{Non-functional oriented adaptivity}: the application is able to execute only one functionality, but with different performance (e.g., the precision of a filter could be reduced to save energy)~\cite{Sau_2017}.
\end{enumerate}

In CERBERO, the Multi-Dataflow Composer (MDC)~\cite{Palumbo2016} tool automates the development of CG-VRCs. Users describe the applications to be accelerated as dataflows and MDC automatically merges them through a datapath merging algorithm, generating a Xilinx-compliant IP with its drivers to delegate computing tasks to the coprocessor~\cite{SauFMRP15}.

%To enable a feedback loop that allows for the design of self-adaptive CPS, the system needs monitors to capture its internal status changes~\cite{Palumbo_2019x2}. 
The first step to enable a feedback loop that allows for the design of self-adaptive CPS, consists of instrumenting the system with monitors to capture its internal status changes~\cite{Palumbo_2019x2}.
The most extended Sw approach for enabling self-awareness is based on accessing the existing Performance Monitoring Counters (PMCs) of modern CPUs. On the other hand, a Hw accelerator can be specialized by the designer to include custom monitors. This second solution is not suitable for Sw developers who may have limited knowledge of the Hw design flow. Furthermore, if these solutions rely on custom methods to read the monitors, the process of reading the monitors in the Hw accelerators and the PMCs already available on the CPU could not be the same, and heterogeneity of solutions, complex to be implemented, may  be required. In CERBERO, \textsc{Papify}~\cite{Madronal_2019_Access,dmadronal2018CF} provides a lightweight monitoring infrastructure by means of an event library aimed at generalizing the Performance Application Programming Interface (PAPI)~\cite{papi} for embedded heterogeneous architectures. 

%The work presented in this paper relies on the idea of offering to Sw developers the support to design and implement run-time reconfigurable systems as the CG-VRCs and, at the same time, to monitor both the processor and the Hw accelerator using a unified methodology based on \textsc{Papify}. 

In a previous work~\cite{Madronal_2019} we proposed the idea of using \textsc{Papify} in combination with MDC to offer support for the design, implementation and monitoring of run-time reconfigurable systems, as the CG-VRCs, using \textsc{Papify}. In that work we presented a PAPI-compliant component that could be automatically configured with events information using an XML file. The work presented in this paper relies on the idea of offering to Sw developers the support to design and implement run-time reconfigurable systems and  to monitor both the processor and the Hw accelerator using a unified methodology based on \textsc{Papify}. Being in a heterogeneous-core computing era, a unified methodology allows a fairer comparison of Hw and Sw performance and facilitates the performance analysis in terms of debugging (e.g., monitor the correct execution of internal modules) and optimization (e.g., monitoring of CG-VRC allows for  prospectively switching  among different configuration if the users require better performance).

\begin{itemize}[leftmargin=*]
\item In this work the MDC tool has been extended to provide automatic instrumentation of the CG-VRCs with custom PMCs and to automatically generate the XML file necessary to automatically configure the previous developed PAPI-component. This automatic flow allows Sw programmers to define the applications to be accelerated and instrumented as dataflow descriptions, without the need of any Hw knowledge.

\item The Application Programming Interfaces (APIs) provided by MDC, in combination with the Sw libraries provided by \textsc{Papify}, offer the transparent PAPI-compliant access to the Hw PMCs. 

\item The monitoring of heterogeneous Hw/Sw systems is a mandatory step to allow self-adaptation of CPS. Nevertheless, in this preliminary exploration the design under test is not a CPS one. Assessment on a processor-coprocessor system for image processing, validates the automatic design flow, the monitoring PAPI-based approach and the effectiveness of \textsc{Papify} on heterogeneous Hw/Sw systems.
\end{itemize}

The paper is organized as follows: Section~\ref{ss:soa} explores the solutions at the state of the art, Section~\ref{s:toolchain} presents the proposed Hw/Sw unified monitoring approach together with the exploited tools, and Section~\ref{s:assessment} presents a proof of concept evaluation of the effectiveness of the approach. At the end, Section~\ref{ss:conc} summarizes and concludes the paper with some directions for future works.

\section{Related Works}
\label{ss:soa}

%SW approaches for monitoring
In literature, several works have dealt with the issue of monitoring Sw and Hw systems to gather relevant data on the system status and its performance for run-time evaluation and/or adaptivity purpose.
In particular, PAPI provides a unified method to access the PMCs available on the CPUs~\cite{papi}.
%copied from the poster
The PAPI community is big, and there are several research works. For instance, Adhianto et al.~\cite{adhianto2010hpctoolkit} proposed a sampling monitoring infrastructure based on PAPI applied to High-Performance Computing systems, and Kn{\"u}pfer et al.~\cite{knupfer2008vampir} focused on providing a graphical interface to analyze trace data based on already completed application executions. 
While \textsc{Papify} generalizes PAPI for embedded heterogeneous architectures~\cite{Madronal_2019_Access}.
%end copied from the poster

% Hw approaches for monitoring
To implement self-aware run-time Hw adaptation, a proper instrumentation of the target substrate with monitors is necessary. Various examples of monitoring solutions are available at the state of the art, for instance, in the AMD64~\cite{amd64} and Intel~\cite{intel} processors. The AMD64~\cite{amd64} presents a Lightweight Profiling (LWP) extension to allow user mode processes to gather run-time performance data %about themselves 
with very low overhead, while the Intel Processor Trace (IPT)~\cite{intel} offers Hw performance counters and Sw able to use information acquired at low-level.

% Hw approaches for monitoring
Other works focus on the custom instrumentation of existing Hw architectures. Schmidt et al.~\cite{Schmidt_2012} proposed Hw Performance Monitoring Interface, which involves the insertion of performance monitoring networks into existing Hw designs, and Patrigeon et al.~\cite{Patrigeon_2018} presented an FPGA-based  platform, instrumented with monitors, for real-time evaluation  of  Ultra Low Power Systems on Chip. While Valente et al.~\cite{Moro_2015, Valente_2016} defined a custom profiling system for embedded applications and a library of elements to compose a Hw profiling system for specific applications. 

% Hw approaches for monitoring for Sw developers
Generally speaking, Sw developers do not have deep knowledge of the Hw design flow and, to ease their access to run-time data, Application Program Interfaces (APIs) should be offered them. APIs are a perfect solution to 
\begin{itemize}[leftmargin=*]
\item hide the details behind the definition and customization of dedicated monitoring infrastructures; and
\item ease the usage of Sw monitoring calls in the application.
\end{itemize} 
The Xilinx SDSoC Development Environment gives users the possibility to use counters in ARM Cortex A9 and performance monitoring units in programmable logic side~\cite{sdsoc}. While Shannon et al.~\cite{Shannon_2015} presented ABACUS, a performance-monitoring framework that can be used to debug the execution behaviours and interactions of multi-application workloads in reconfigurable logic scenarios. 

% Hw approaches for monitoring using papi
Some works have tried to exploit PAPI for Hw systems. For instance, Ho et al.~\cite{Ho_2014} proposed a performance monitoring unit integrated with the $perf\_event$ API. Suriano et al.~\cite{Suriano_2018} presented a custom approach that uses \textsc{Papify} for reading monitors of a Hw slot-based architecture that exploits Dynamic and Partial Reconfiguration~\cite{Rodriguez2018}. 

With respect to above works, this paper proposes a generic PAPI-based approach for monitoring Hw accelerators, suitable also to enable the proper feedback of a CG-VCR architecture. In particular, we extended the MDC tool to automatically instrument with custom PMCs the generated Hw accelerators, providing the necessary Sw support to monitor both Sw and Hw Processing Element (PEs) through the same interface, \textsc{Papify}.

\section{Toolchain for Heterogeneous Monitoring}
\label{s:toolchain}

This section presents the toolchain for the development, implementation and management of monitored heterogeneous platforms. Section~\ref{ss:mdc} depicts the design flow of MDC tool, Section~\ref{ss:papify} illustrates the run-time monitoring capabilities of the \textsc{Papify} Tool, and Section~\ref{ss:hw-monitoring} presents the proposed monitoring approach. Indeed, a set of definitions must be clarified in advance:
\begin{itemize}[leftmargin=*]
    \item Dataflow (DF): an application represented as a set of functional elements, \emph{the actors}, exchanging data, the \emph{tokens}, through a set of communication links, the \emph{edges}.
    \item Actor: univocal functional element, encapsulating a given functionality or operation, in which a dataflow application is divided. 
    \item Edge: exclusive interconnection between two actors implemented as a FIFO.
    \item Processing Element (PE): Hw resource where one or more actors are scheduled for execution. It can be a Sw core or a complete Hw accelerator.
    \item  Functional Unit (FU): custom  implementation of one single actor instance inside the Hw accelerator. FUs can be manually defined or synthesized using High Level Synthesis tools.
\end{itemize}

\subsection{The Multi-Dataflow Composer Tool}
\label{ss:mdc}

\begin{figure}[h!t]
	\centerline{\includegraphics[width=0.49\textwidth]{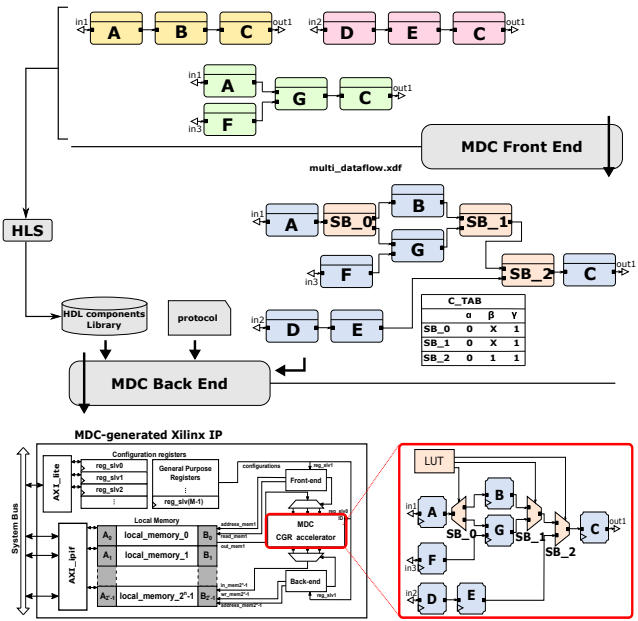}}
	\caption{MDC design flow.}
	\label{fig:mdc-flow}
\end{figure}

\begin{figure*}[h!t]
	\centerline{\includegraphics[width=0.9\textwidth]{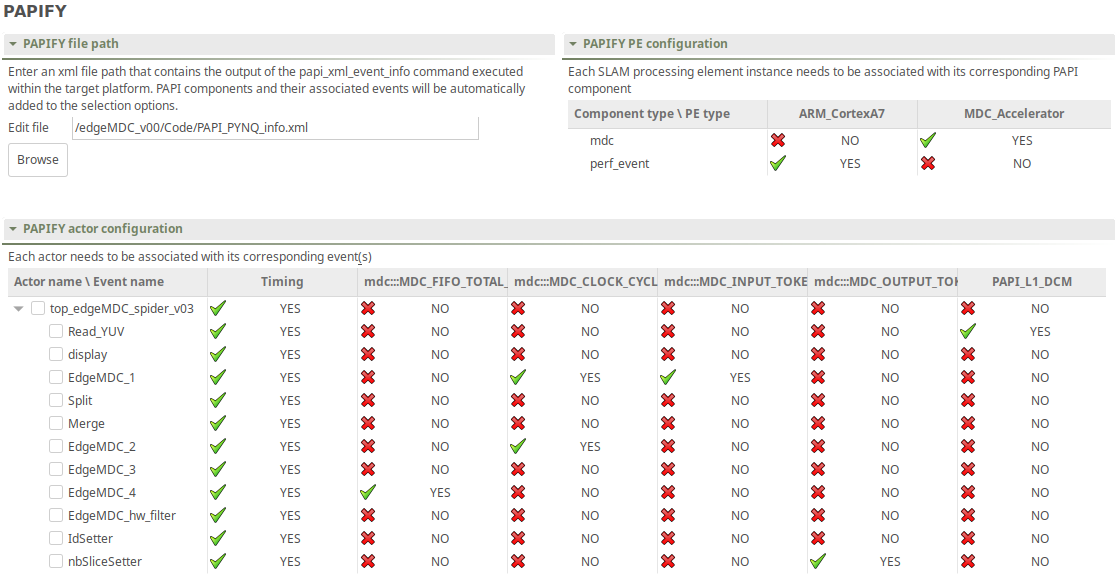}}
	\caption{\textsc{Papify} configuration using PREESM.}
	\label{fig:papifyPreesm}
\end{figure*}

A Hw design can be described as a modular composition of FUs. The same level of expressiveness can be given by a higher level representation, as a dataflow network, where each FU can be represented by an \textit{Actor} of the network, through a $1:1$ mapping.  Several high level description, mapped on one unique Hw description, can represent a CG-VRC design. However, the mapping is not longer $1:1$, but it becomes $N:1$, through the sharing of the common \textit{Actor} and the insertion of ad hoc switching elements.

The Multi-Dataflow Composer (MDC) is an automated framework that generates heterogeneous and irregular CG-VRCs, through an application-to-hardware approach. Applications to be implemented are specified as XML Dataflow Format (XDF)  models and combined through a datapath-merging algorithm that merges the input specifications and allows sharing the actors in common among the different dataflow applications. In the resulting multi-functional CG-VRC heterogeneous accelerator, in Verilog Hardware Description Language (HDL), the FUs are actor-specific. To access shared resources,  multiplexers named Switching-Boxes (\textit{SBoxes}) are inserted in the datapath. The user is required to model the applications to be accelerated as dataflows while MDC takes care of automatically generating the corresponding CG-VRC accelerator.

Top part of Figure~\ref{fig:mdc-flow} depicts an example of the MDC operation. Nodes of the networks (i.e., \textit{A}, \textit{B}, \textit{C}, etc\dots) are \textit{Actors}. The three input dataflow specifications are merged into a multi-functional dataflow, in which the switching elements (e.g., \textit{$SB\_0$}, \textit{$SB\_1$}, \textit{$SB\_2$}) guarantee the correct operation of the different functionalities. During the merging process MDC keeps also trace of the programmability of the switching elements ($C\_TAB$). The Hw description of the single \textit{Actors} (\textit{HDL Component Library}) can be manually written or automatically generated by means of High Level Synthesis (HLS) tools, as for instance CAPH~\cite{serot2013caph}. MDC properly connects them keeping into account handshake protocol among the \textit{FUs} thanks to the communication protocol specified as input file (\textit{protocol}).

MDC also offers the possibility of seamlessly integrating the CG-VRC logic into a processor-coprocessor system for Xilinx environments~\cite{SauFMRP15}. By analyzing the features of the combined dataflow specifications, suitable wrappers for different processor-coprocessor communication infrastructures (memory-mapped or stream) are automatically provided.  The bottom part of Figure~\ref{fig:mdc-flow} shows the resulting CG-VRC embedded in the Xilinx IP generated by MDC.

MDC also provides the APIs to  delegate computation to the coprocessing unit and manage processor-coprocessor communication, masking the system configuration complexity, providing a C function for each configuration of the CG-VRC coprocessor that allows the user to access the accelerator transparently, without taking care of the implementation of data transmission according to the implemented bus protocol. The Listing~\ref{list:int_dr} shows the \textit{C} interface for one configuration of a memory-mapped CG-VRC coprocessor computing the \textit{Roberts} edge detection algorithm. \texttt{data\_<port\_name>} and \texttt{size\_<port\_name>} are respectively input (or output) port and the number of data related to that port. In the considered example there are three ports: \texttt{in\_size}, \texttt{in\_data} and \texttt{out\_data}.

\lstset{language=C,caption={Coprocessor drivers interface.}, commentstyle=\color{mygreen}, keywordstyle=\color{mymauve}, label={list:int_dr},basicstyle=\footnotesize}
\begin{lstlisting}
//Memory-Mapped Interface Driver
int mm_accelerator_roberts(
// port out_data
int size_out_data, int* data_out_data,
// port in_data
int size_in_data, int* data_in_data,
// port in_size
int size_in_size, int* data_in_size);
\end{lstlisting}

The underline \textit{C} code manages the co-processor configuration and data transfer. For each I/O port of the reconfigurable computing core, a configuration word (\texttt{size\_in\_size}, size\_in\_data, size\_out\_data) is written into the proper co-processor register. Then, for each input port involved in the current computation, a specific primitive is used to send the data (\texttt{data\_in\_size}, \texttt{data\_in\_data}) to be computed from the host processor to the co-processor. At last, a specific primitive is adopted to read back the results (\texttt{data\_out\_data}) into the processor from the output ports.

\subsection{The \textsc{Papify} Tool}
\label{ss:papify}

\textsc{Papify}~\cite{Madronal_2019_Access} is a tool aiming at easing the instrumentation and PAPI-based monitoring of applications. In order to use it, a dedicated library called \emph{eventLib} is available.
This library is composed of 9 functions and is built on top of PAPI. 
With this library, the user only needs to include a set of functions at the beginning of the application where all the monitoring is configured. Additionally, both the PE (i.e., a physical Sw core or an accelerator) and the actor (functional block) configurations are isolated from each other. During the configuration of the PE monitoring, the available PAPI components are linked to the corresponding PEs. Secondly, the configuration of the actors is performed associating events contained in any PAPI component to it.
%{\color{blue} TF: I think it's important to specify what's PE and what's actor (I guess that presenting also the workflow with preesm and spider would help in this.)}{\color{purple} DM: I added the (functional unit) for the actor because it is the first reference, but during the introduction, the concept of PE is explained. I will add at the end of the section the use of PAPIFY within a dataflow approach (SPiDER) and everything will be clear then.)}

Once the monitoring configuration is included, the user only needs to set the starting and stopping points for the monitoring. By doing so, the instrumentation of the code will be complete and, independently of the PAPI component that is accessed and the PE that is executing the actors, the structure of the monitoring will be homogeneous. 

Regarding the behaviour in execution time, \textsc{Papify} manages the different configurations for PEs and actors combining them in a completely transparent way to the user. That is, it automatically selects the PAPI events that are available for the specific PE and stores the results accordingly. Additionally, \textsc{Papify} stores the configurations that have been already set up during one execution in order to reuse it, hence, reducing the monitoring overhead.

Listing~\ref{list:papifyUsage} depicts an example of code necessary to instrument an application using \textsc{Papify}. The user can monitor the application instrumenting the code by hand or automatically, thanks to the integration of \textsc{Papify} with a dataflow development framework called PREESM~\cite{Pelcat2014}.
%Even if the user can monitor applications using \textsc{Papify} by hand (as shown in Listing~\ref{list:papifyUsage}), it has been integrated with a dataflow development framework called PREESM~\cite{Pelcat2014}. 
%This framework 
PREESM provides automatic code generation of dataflow applications and, together with \textsc{Papify}, automatic code instrumentation is provided to the user. The user only needs to fill a graphical configuration as the one shown in Figure~\ref{fig:papifyPreesm}, following the same organization shown in Listing~\ref{list:papifyUsage}.

\lstset{language=C,caption={PAPIFY usage example.}, commentstyle=\color{mygreen}, keywordstyle=\color{mymauve}, label={list:papifyUsage},basicstyle=\footnotesize}
\begin{lstlisting}
//Initial monitoring configuration
configure_papify_PE(char* coreName,
char* PAPIComp,int PEid);
configure_papify_actor(&papify_action,
char* PAPIComps,char* actorName,int numOfEvents,
char* eventNames,char* configIDs,int numConfigs);
// Monitoring
event_start(&papify_action,int PEid);
actor_to_be_monitored();
event_stop(&papify_action,int PEid);
\end{lstlisting}

%Specifically, the 
The run-time use of \textsc{Papify} in the context of reconfigurable dataflow, as the one of the SPIDER tool~\cite{Heulot2014}, is graphically explained in Figure~\ref{fig:papifySpiderFlow}. As can be seen, the process is divided into five steps:

\begin{figure}[h!t]
	\centerline{\includegraphics[width=0.49\textwidth]{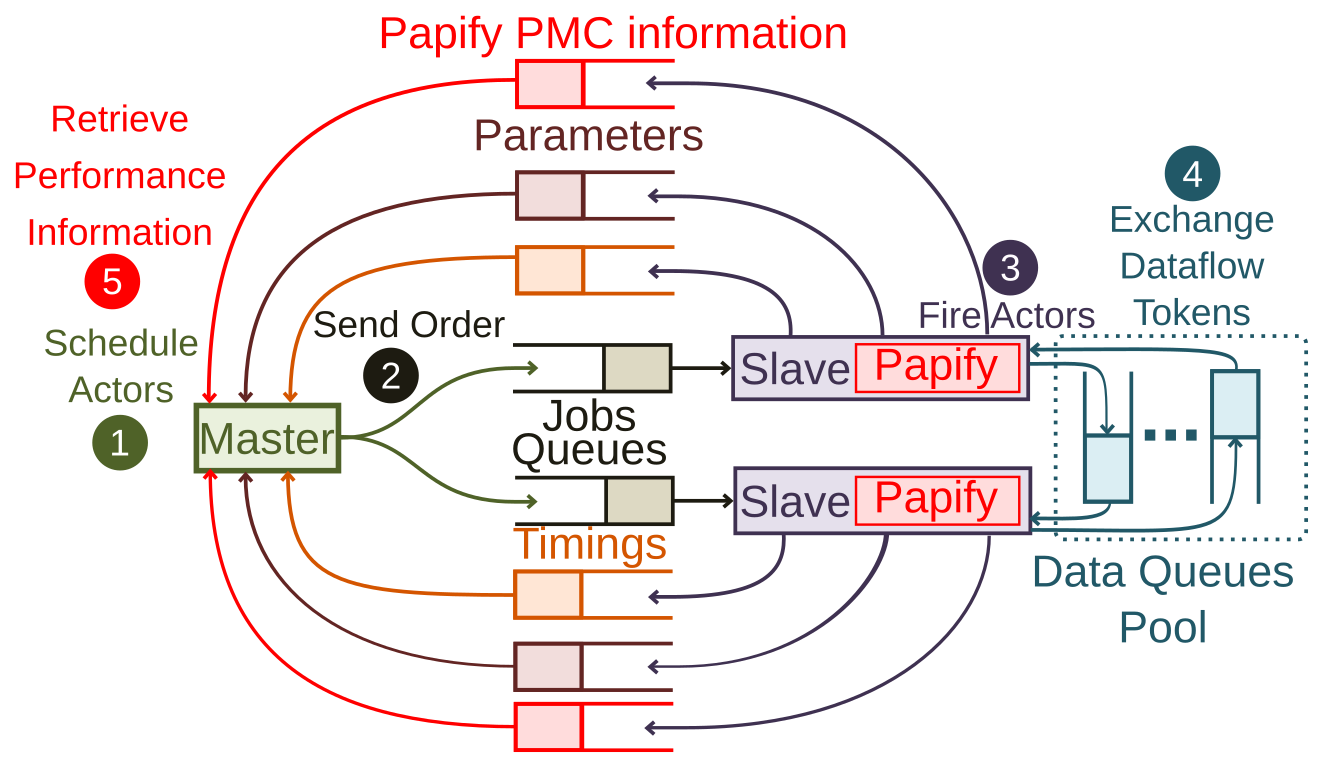}}
	\caption{\textsc{Papify} in a dataflow context.}
	\label{fig:papifySpiderFlow}
\end{figure}

\begin{enumerate}[leftmargin=*]
\item \textit{Schedule Actors}: a master process is in charge of scheduling the actors composing the dataflow application.
\item \textit{Send Order}: the master process maps the application actors over the available slave PEs (either Sw cores or Hw accelerators).
\item \textit{Fire Actors}: PEs execute the scheduled actors and, during these executions, the PAPI events are retrieved using \textsc{Papify}.
\item \textit{Exchange Dataflow Tokens}: PEs, according to the application flow, exchange tokens. 
\item \textit{Retrieve Performance Information}: Once the whole application has been executed, the performance data is retrieved by the master process together with timings and application parameters. This will enable the master process to take re-mapping and scheduling decisions based on this new information. 
\end{enumerate}

\subsection{Proposed approach}
\label{ss:hw-monitoring}

The proposed approach relies on the integration of \textsc{Papify} and MDC, to provide a toolchain able to offer the support in the process of designing, implementing and managing monitored CG-VRCs. \textsc{Papify} provides an interface to access performance monitoring information of the different PEs existing in the target platform. As PAPI is built based on components, (i.e., each resource is isolated from each other to separate also the low-level details of each Hw resource) \textsc{Papify} automatically inherits this structure. Nevertheless, this new tool has been built to transparently manage the monitoring configuration independently of the nature of the PE executing each part of the application. 

\begin{figure}[h!t]
	\centerline{\includegraphics[width=0.45\textwidth]{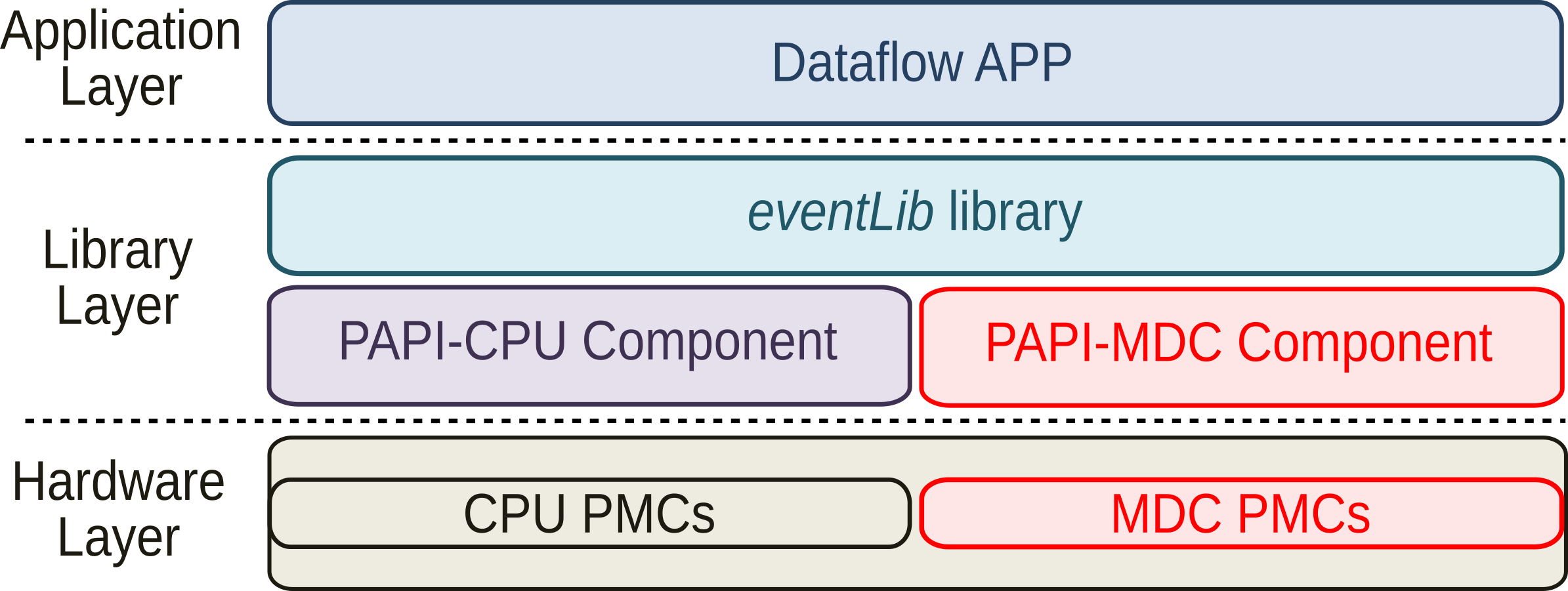}}
	\caption{Hw/Sw monitoring infrastructure, with custom PAPI component and PMCs for MDC.}
	\label{fig:papi-component}
\end{figure}

Figure~\ref{fig:papi-component} illustrates the PAPI-based monitoring infrastructure. At the \textit{Application Layer} the user only needs to specify the application as dataflow and instrument it with the \textsc{Papify} calls.
On the bottom, the \textit{Hardware Layer} presents the PMCs for both the host processor (CPU PMCs)  and the Hw accelerators on the FPGA (MDC PMCs). At the \textit{Library Layer}, the PAPI components are the \textit{C} interface between the high-level user \textit{Application Layer} and the  \textit{Hardware Layer} and take care of accessing the PMCs. To eases both the configuration and the management of the monitoring, including the transparent access to the PAPI components associated with either Sw or Hw PEs, \textsc{Papify} offers the \textit{eventLib} Library.
Enabling the monitoring of the Hw CG-VRCs using \textsc{Papify}, required to develop both a PAPI component and the PMCs suitable for MDC.
% end copied from the poster

\begin{figure}[h!t]
	\centerline{\includegraphics[width=0.49\textwidth]{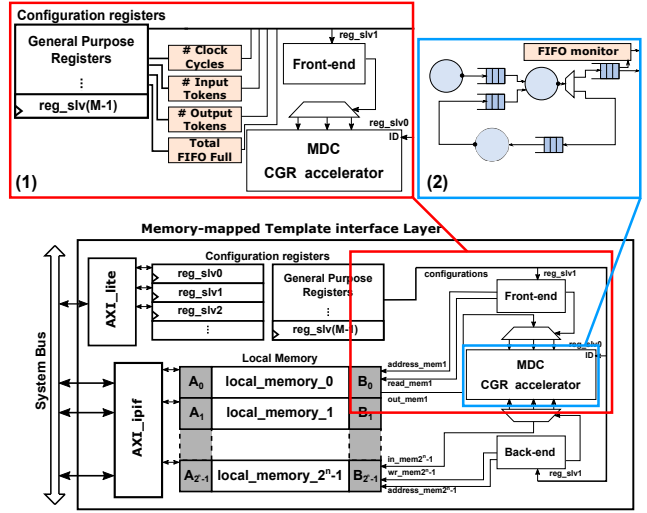}}
	\caption{Hw Monitoring at two levels of abstraction: (1) accelerator level; (2) low-level.}
	\label{fig:Hw-monitors}
\end{figure}

As mentioned in Section~\ref{s:intro}, in modern CPUs there are built-in PMCs to monitor various kinds of events, while on Hw accelerators it is necessary to rely on custom solutions. 
% copied from the poster
Figure~\ref{fig:Hw-monitors} illustrates an MDC-generated  IP, instrumented with monitors. This memory-mapped IP communicates with the host processor (not shown in Figure~\ref{fig:Hw-monitors}) through the \textit{System Bus}. The \textit{MDC CGR Accelerator} is connected to the \textit{System Bus} by means of  a memory bank (\textit{Local Memory}) to exchange data between the processor and the coprocessor, and through a register bank (\textit{Configuration Registers}) to send the words necessary to configure the coprocessor. In this IP the monitors are placed at two levels of abstraction: 

\begin{enumerate}[leftmargin=*]
    \item accelerator-level: this monitoring, placed outside the \textit{MDC CGR Accelerator}, is homogeneous for every accelerator that can be implemented using MDC. It keeps trace of standard dataflow metrics during execution, such as the execution time, the number of input tokens and the number of output tokens. 
    \item low-level: this monitoring, placed inside the \textit{MDC CGR Accelerator}, is specific for the current accelerator, e.g. by profiling the bottleneck FUs internally. 
\end{enumerate}

 The accelerator-level monitors are automatically inserted by MDC, while at the moment the low-level monitoring still requires manual steps to be used within the IP. In both cases, the HW monitors are accessed by the host processor through the \textit{Configuration Registers} of the IP.
 
Since the base address of the accelerator may change in different accelerators, as well as the number and type of events to be monitored, we developed a configurable PAPI-MDC component that is automatically configured when the application is launched. The PAPI-MDC component is compliant with the existing Sw component standard, and can be naturally accessed by \textsc{Papify}. When the application is launched, and the specific monitors for the accelerator under evaluation are loaded, the PAPI-MDC component is automatically configured, using a configuration XML file, as the one depicted in Listing~\ref{list:conf_file}. 
In the current XML file the user shall specify the physical base address of the accelerator to be monitored (\texttt{baseAddress}), the number of available events (\texttt{nbEvents}) and their type (\texttt{event}), but this approach can be easily extended to consider other variables. This kind of monitoring is transparent for the user, that only needs to insert the \textsc{Papify} call in the Sw application. 

\lstset{language=XML,caption={Configuration file for the PAPI component.}, morekeywords={mdcInfo,baseAddress,nbEvents,event, index,name,desc}, commentstyle=\color{mygreen}, keywordstyle=\color{mygreen}, label={list:conf_file},basicstyle=\footnotesize}
\begin{lstlisting}
<mdcInfo>
  <baseAddress>0xADDRESS</baseAddress>
    <nbEvents>N</nbEvents>
    <event> 
      <index>M</index>
      <name>MDC_EVENT_NAME</name>
      <desc>Event Description</desc>
    </event>
</mdcInfo>
\end{lstlisting}

The heterogeneous Hw/Sw system that can be monitored with such approach is a SoC platform like the ones provided by the Xilinx Zynq family, in which the chip includes both one or more processors and a programmable logic part. The board requires to run Linux, in which are installed PAPIFY and PAPI (that needs to include the PAPI-compliant MDC-component).

The Hw accelerator is modelled starting from a set of XDF dataflow networks that are parsed by MDC to generate a Xilinx-compliant IP able to execute all the different functionalities described by the input dataflow specifications, one at a time (see Section~\ref{ss:mdc}). MDC also generates the \textit{C} APIs to mask the communication between the processor and the coprocessor. Thanks to the extension of MDC, the user can specify to instrument the generated Verilog HDL code to  incorporate the accelerator-level monitors above described and to generate the XML file necessary to configure the PAPI-compliant MDC-component.

This Hw accelerator is used by a Sw \textit{C/C++} application, that can be either manually developed or modelled as a PiSDF specification using the \textsc{Papify}-PREESM-SPiDER flow described in Section~\ref{ss:papify} that automatically integrates the \textsc{Papify} monitoring code.

When the application is launched the PAPI-MDC component is automatically configured using the generated configuration XML file described above. The PAPI components access the PMCs and \textsc{Papify} collects the data and save them into csv files that can be analyzed using \textsc{Papify-Viewer}.

\section{Assessment}
\label{s:assessment}

In this Section, as a proof of concept, the proposed monitoring approach and toolchain is evaluated through an application for Image Processing, involving a multifunctional accelerator for edge detection, able to compute two different algorithms: \textit{Sobel} and \textit{Roberts}. The system is based on the Zynq-7000 XC7Z020CLG484-1 device running Linux. It is composed of a Sw Application, mapped among the two ARM Cortex A9 cores available on the adopted board, that acquires an input video (Section~\ref{ss:sw}) and of a Hw accelerator, implemented on the Programmable Logic of the FPGA, to compute the edge detection (Section~\ref{ss:hw}). 

\subsection{Design Under Test}
\label{ss:dut}

\begin{figure}[h!t]
\centering
\includegraphics[width=0.49\textwidth]{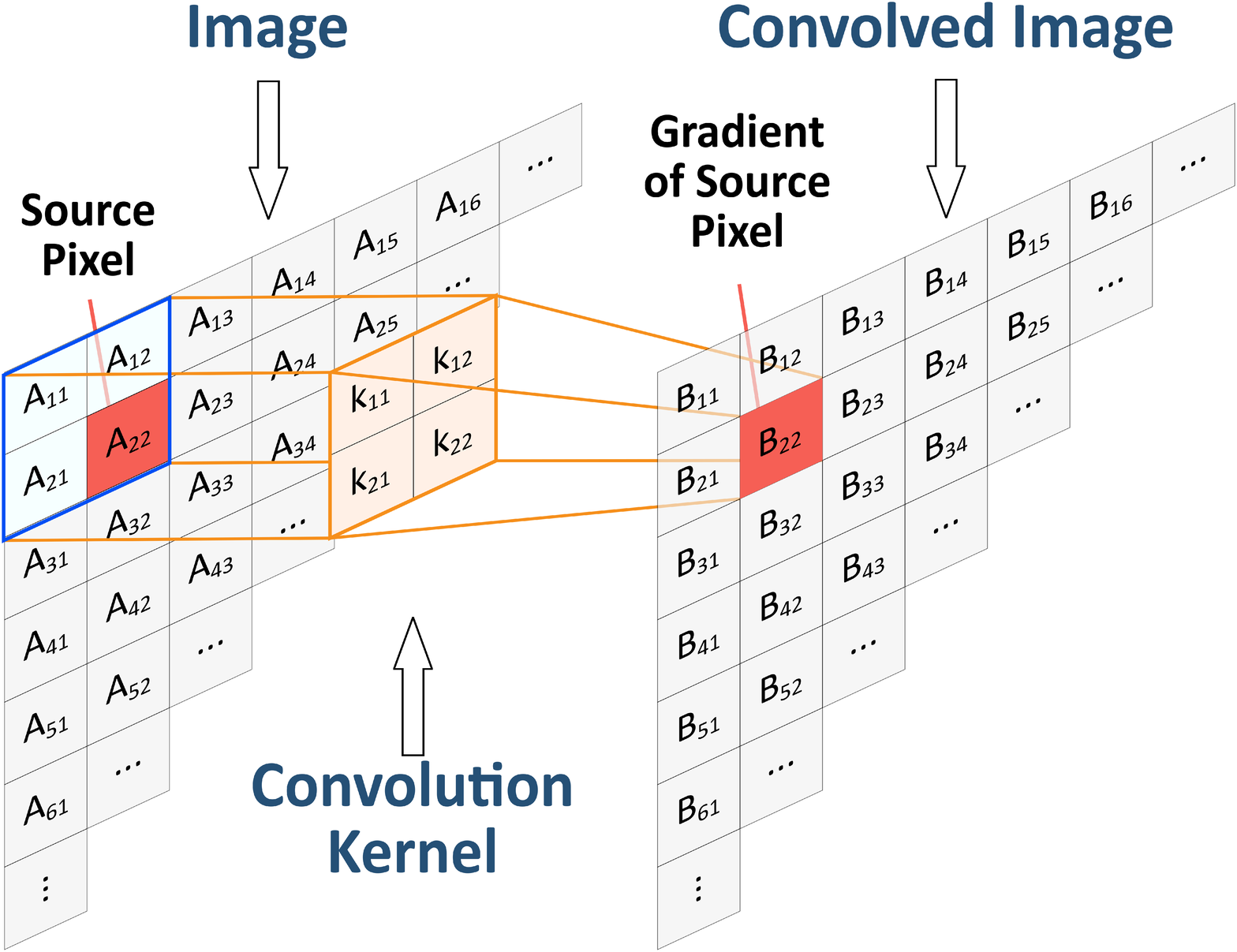}
\caption{Overview of the convolution operation in \textit{Roberts} edge detectors.}
\label{fig:roberts_conv}
\end{figure}

Edge detection algorithms estimate the magnitude and the orientation of edges on an image~\cite{Davies84}. In this proof of concept we adopted discrete first-order differentiation operators, in which the boundary of an object is the difference of the intensity levels in its pixels with respect to the surrounding pixels. These operators are applied to evaluate the gradient image ($G$), that corresponds to the magnitude of the edge. The computation consists of a convolution of a 3x3 kernel ($k$) with the source image ($A$) for \textit{Sobel}, while for \textit{Roberts} $k$ is 2x2: $G = k * A$. Figure~\ref{fig:roberts_conv} graphically illustrates the convolution operation necessary to compute \textit{Roberts} algorithm. For a complete description of the adopted Edge Detection algorithms please see~\cite{fanni_2018}.

Both the Sw and the Hw parts of the system are modelled as dataflow networks, and are described more in details in the Section~\ref{ss:sw} and Section~\ref{ss:hw} .

\subsubsection{Sw Application}
\label{ss:sw}

\begin{figure*}[h!t]
	\centerline{\includegraphics[width=0.99\textwidth]{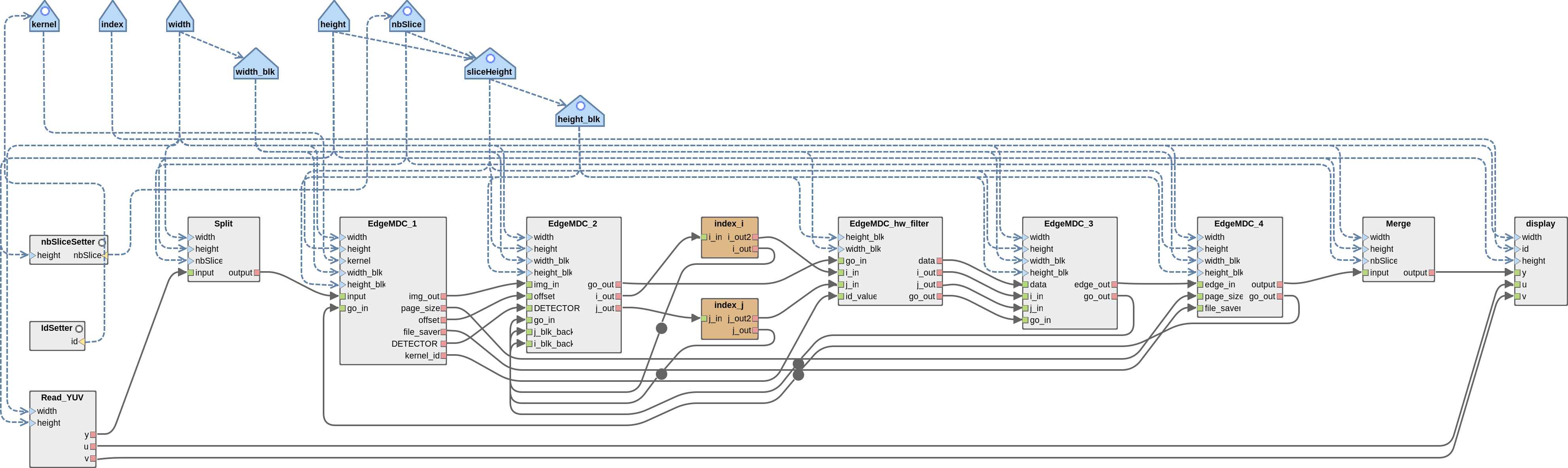}}
	\caption{Dataflow description of the Sw application, modelled using PREESM~\cite{preesm}. Actors and routing blocks are respectively represented by grey and orange boxes. Blue pentagons correspond to the PiSDF parameters that, given as inputs to the actors, establish their specific functionalities. Connections among blocks depict the data token transfer links (grey wires) and the dependencies from the parameters (blue dashed lines).}
	\label{fig:top-spider}
\end{figure*}

The Sw application has been designed using a Parameterized  Interfaced Synchronous Dataflow (PiSDF) specification and the toolchain composed of the design-time tool PREESM~\cite{Pelcat2014} and the run-time manager SPIDER (integrating the automatic code generation of \textsc{Papify} monitoring). In this context, actors exchange tokens through edges depending on the feasible working points of the application scenario. %functional blocks (actors) exchange data through communication links (edges) depending on the feasible working points of the application scenario. 
The configuration parameters established at design-time and run-time are respectively called \textit{static} and \textit{dynamic}. The latter ones imply on-the-fly re-scheduling and re-mapping when their values change. 
The use-case algorithm depicted in Figure~\ref{fig:top-spider} can be described as follows:
\begin{itemize}[leftmargin=*]
    \item Given as an input to the actor \textit{Read\_YUV}, a YUV video is read frame by frame, where the number of rows and columns correspond to $height$ and $width$ parameters respectively. Filtering is applied only to the Y component, while the other ones are directly sent to be displayed. 
    \item Before the edge detection, the block \textit{Split} divides the image in slices depending on the degree of exploitable parallelism. In this assessment, having available one single Hw accelerator, no adaptation has been considered in this sense ($nbSlice = 1$, that is $sliceHeight=height$).
    \item At this point, verified the on-the-fly selected \textit{kernel} (set by \textit{IdSetter}) among $Sobel$ and $Roberts$, an initialization phase is performed %as described 
    in \textit{EdgeMDC\_1}. In this phase, the processing data and the communication with the %reconfigurable device 
    accelerator (through the Direct Memory Access) are handled.
    \item Then, processing occurs by blocks of pixels of a size suitable for the accelerator specifications (in the assessed example, $32\times32$). \textit{EdgeMDC\_2} sends a number of blocks corresponding to $width\_blk\times height\_blk$ to the \textit{EdgeMDC\_hw\_filter}, which forwards the data to the coprocessor. Therefore, \textit{EdgeMDC\_3} receives the result of each iteration, which is collected in \textit{EdgeMDC\_4}.  
    \item Finally, the filtered frame is merged and displayed with the applied type of kernel and the execution time expressed in Frames per Second (FpS).
\end{itemize}
With respect to the mapping strategy, SPIDER handles all Sw tasks taking into account the constraints given as input by the application designer. In the evaluated case, the actors performing splitting and merging have to be executed onto the same core. Moreover, SPIDER has managed 305 instances of the single-rate graph. Indeed, 8 actors are executed 1 time per firing, and 99 times the other 3 ones (\textit{EdgeMDC\_2},
\textit{EdgeMDC\_hw\_filter}, and \textit{EdgeMDC\_3}), since 99 $32\times32$ blocks are present in the frame size considered in this assessment (352x288 pixels). Regarding the actual filtering, this has been accelerated on Hw, as explained in Section~\ref{ss:hw}.

\subsubsection{Hw accelerator}
\label{ss:hw}

\begin{figure}[h!t]
\centering
\includegraphics[width=0.5\textwidth]{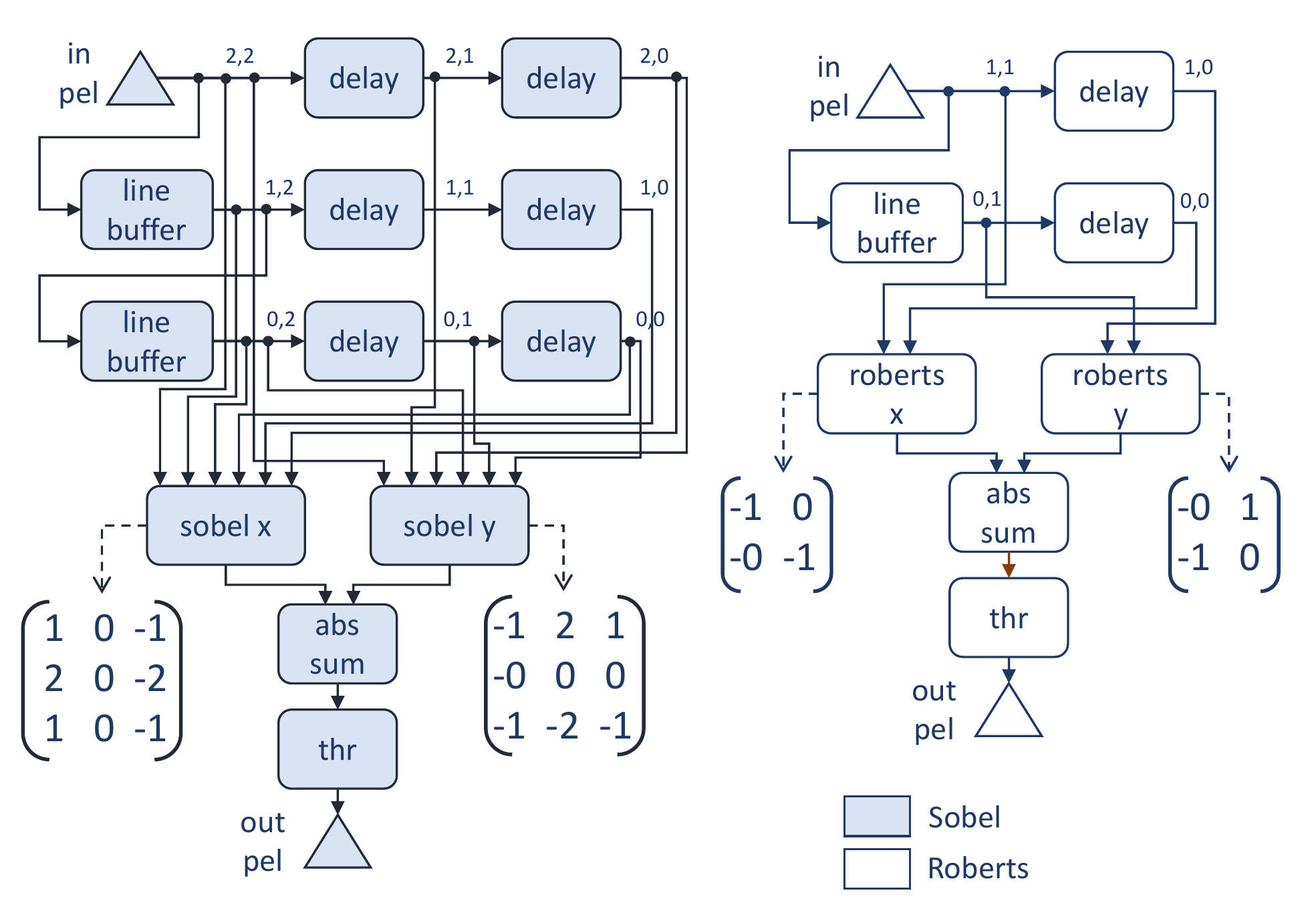}
\caption{Schematic graphs representation of the \textit{Sobel} and the \textit{Roberts} edge detectors.}
\label{fig:graph}
\end{figure}

As described in Section~\ref{ss:mdc}, MDC takes as input the dataflow descriptions of the applications to be accelerated. In this work the dataflow processed by MDC have been described in CAPH language~\cite{serot2013caph}, considering, to implement the CG-VRC accelerator, the the design flow proposed in~\cite{Rubattu2018}, which enables benefits compared to the main tools available in the market (e.g., Intel FPGA SDK for OpenCL~\cite{opencl} and Xilinx Vivado HLS~\cite{vivadohls}).

Figure~\ref{fig:graph} depicts a schematic graph representation of the Sobel and Roberts kernels. The \textit{line buffer} actors are adopted to store previous rows of the image, while \emph{delay} actors are in charge of memorizing one previous pixel within a row. Once the actors are filled with the proper numbers of rows and pixels, the convolution actors can compute the horizontal and vertical gradients. Actor \textit{abs sum} sums up the absolute values of the horizontal and vertical gradients and right-shifts the result for a given scaling factor $n$. Lastly, the thresholding actor \textit{thr} sets to 255 all the magnitudes that are above a the threshold (in this case it has been fixed to 80), while setting to 0 the others.

These dataflow specifications has been processed by MDC, to generate a CG-VRC able to compute both \textit{Sobel} and \textit{Roberts} algorithms, which has been automatically embedded into the ready-to-use Xilinx IP.

\subsection{Experimental Results}
\label{ss:results}

The described Sw application has been mapped onto two cores. Specifically, \textit{display} and \textit{Read\_YUV} actors are mapped onto the \textit{Core 0} while the others are mapped onto the \textit{Core 1} of the adopted board. Among the actors mapped onto \textit{Core 1} three actors are repeated more than one time per firing: \textit{EdgeMDC\_2}, \textit{EdgeMDC\_hw\_filter}, and \textit{EdgeMDC\_3}. 
These actors are executed for each $32\times32$ block of the frame (in our case we have 99 blocks). As explained in Section~\ref{ss:sw} the \textit{EdgeMDC\_hw\_filter} takes care of communicating with the Hw accelerator to compute the edge detection.
On the bases of the described design under test, three different configurations are evaluated.

\begin{itemize}[leftmargin=*]
    \item \textit{DUT\_1} - Hw/Sw system where the Hw accelerator includes the PMC, and both the Sw application and the Hw accelerator are monitored. In the Sw application the \textit{display} and \textit{Read\_YUV} actors are selected for the monitoring of the clock cycles and number of instructions events, while in the Hw accelerator the monitored events are the execution time (clock cycles) and the throughput (number of output tokens).
    \item \textit{DUT\_2} - Hw/Sw system in which the Hw accelerator includes the PMC, but no monitoring is performed. 
    \item \textit{DUT\_3} - The same than DUT\_2 but without any PMC inside the Hw accelerator. 
\end{itemize}

Table~\ref{tab:overhead} reports, for \textit{Roberts} execution, the performance of the different designs in terms of average (\textit{FpS}). As expected the monitoring does not come for free, and the monitored design (\textit{DUT\_1}) has a performance loss of 6.20\%, with respect to its no-monitored version (\textit{DUT\_2}). Specifically, this overhead is due to the fact that the application is executing a total amount of 305 actors in each iteration. Consequently, the same amount of lines are written in csv files that are provided to 1) analyze the application and 2) locate possible bottlenecks using \textsc{Papify-Viewer}. Please, note that \textit{DUT\_3} does not have any PMC, thus Table~\ref{tab:overhead} does not report any overhead data.

For the sake of completeness, Table~\ref{tab:overhead} also depicts the performance variation of the not monitored design, in which the accelerator embeds the custom PMCs (\textit{DUT\_2}), with respect to its equivalent version in which the Hw accelerator has not any monitor. In this case, the different performance (1.49\%) are due not to Sw reasons, rather to the different Hw designs that, having a different number of configuration registers and different logic, can be synthesized in a different manner by Xilinx Vivado, leading to different performance in terms of access to the memory and the registers normally used to communicate with the accelerator.

\begin{table}[h!]
    \caption{Performance of the three considered designs. \textit{FpS} is the average frames per second processed by the design. \textit{St. Dev} is the standard deviation of the values.   \textit{Overhead} is the \% performance variation of \textit{DUT\_1} (\textsuperscript{*}wrt \textit{DUT\_2}) and of \textit{DUT\_2} (\textsuperscript{**}wrt \textit{DUT\_1})}
    \begin{center}
        \begin{tabular}{r|c|c|c}
            \textit{Design} & \textit{FpS} & \textit{St. Dev} & \textit{Overhead}  \\
            \hline
            \textit{DUT\_1} &  11.03 & 0.0489 & 6.20\%\textsuperscript{*} \\
            \textit{DUT\_2} &  11.76  & 0.0138 & 1.49\%\textsuperscript{**} \\
            \textit{DUT\_3} &  11.94  & 0.0034 & --- \\
        \end{tabular}
    \end{center}
    \label{tab:overhead}
\end{table}

\begin{figure*}[h!t]
\centering  
\begin{subfigure}[b]{0.99\textwidth}
\centering  
\includegraphics[width=0.99\textwidth]{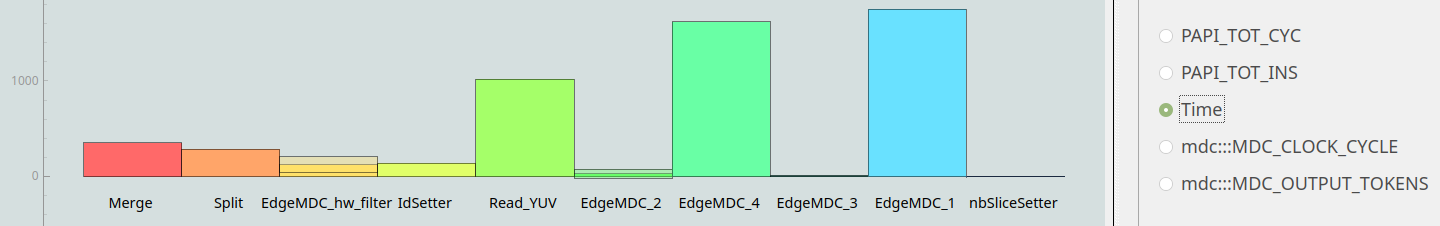}
\caption{Timing event}
\label{fig:timing}
\centering
\includegraphics[width=0.99\textwidth]{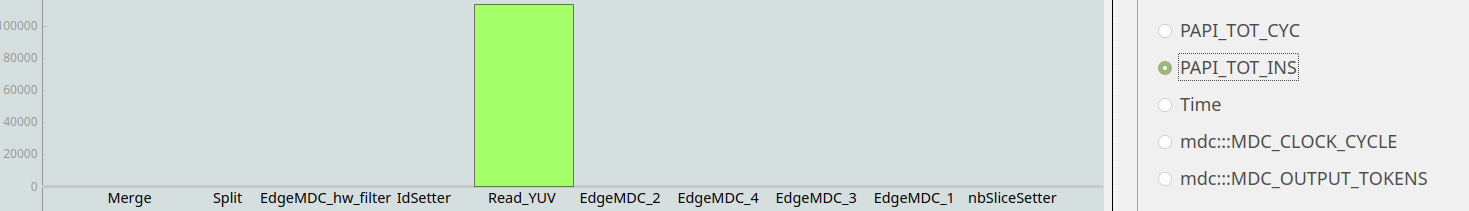}
\caption{$PAPI\_TOT\_INS$ Sw event}
\label{fig:yuv_event}
\centering
\includegraphics[width=0.99\textwidth]{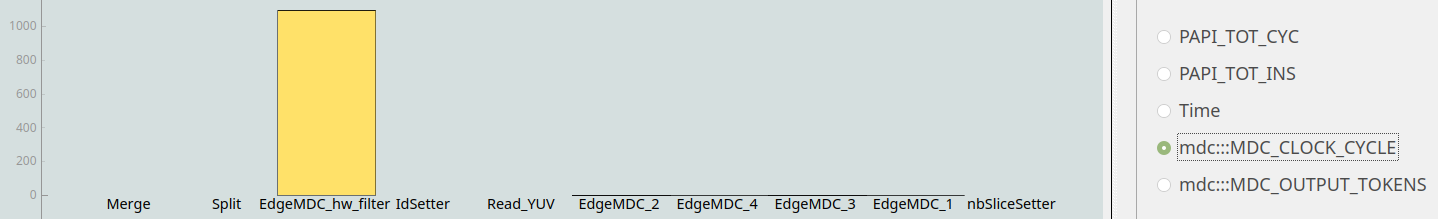}
\caption{$MDC\_CLOCK\_CYCLE$ Hw event}
\label{fig:hw_event}
\end{subfigure}
\caption{Papify-Viewer event plots}
\label{fig:papify_viewer}
\end{figure*}

Finally, the events obtained through \textsc{Papify} can be easily analyzed using its viewer, the so-called \textsc{Papify-Viewer}. As can be seen in Figure~\ref{fig:papify_viewer}, the events obtained at run-time during the execution of Sobel-Roberts application can be analyzed one by one. First, in Figure~\ref{fig:timing}, timing is monitored for every actor and, as can be seen, $EdgeMDC\_1$, $EdgeMDC\_4$ and $Read\_YUV$ are the actors taking longer. This is coherent with the reality because these three actors are the ones managing the whole frame. On the contrary, the actor being executed 99 times per iteration ($EdgeMDC\_hw\_filter$), is one of the fastest actors in the specification. Secondly, in Figure~\ref{fig:yuv_event} and Figure~\ref{fig:hw_event}, events associated to $perf\_event$ and $MDC$ PAPI components are shown, respectively. In here, it can be observed that the events associated to the real execution of the Hw accelerator ($EdgeMDC\_hw\_filter$) are properly measured for the only actor associated to real Hw accelerator execution. 

\subsection{Advantage of the Proposed Approach}
\label{evaluation}

It is important to highlight the effectiveness of the proposed flow in terms of design time and effort.
The design of CG-VRCs requires to analyze the common resources of different dataflow specification, and to combine them, keeping trace of the \textit{Actors} belonging to different functionalities and to program the multiplexers properly. Therefore, the manual design of CG-VRCs is time consuming and error prone.
The proposed approach speeds-up and simplifies the design of monitored CG-VRCs by automatically mapping different input specifications in one MDC multi-flow IP, instrumented with monitors. The usage of dataflow specifications allows for the exploitation of HLS dataflow-to-hardware tools (such as CAPH~\cite{serot2013caph}), which not only speed the design process up by automating HDL generation, but also allow developers that are not expert in Hw design to adopt the proposed approach.
The users only need to define the applications through abstract high level input dataflow specifications; then, the toolchain takes care of the complete process from dataflow to the processor-coprocessor system. The generated APIs mask the complexity of the processor-coprocessor communication, and thanks to Sw-compliant MDC-PAPI component and the support for heterogeneous architectures provided by \textsc{Papify}, the users can transparently access every monitor, by means of a call to a \textsc{Papify} function. \textsc{Papify} autonomously collect the data related to the monitored events and save them in  csv files, offering to the user to post process them and to analyze them using \textsc{Papify-Viewer}.

\section{Conclusions and Future Works}
\label{ss:conc}

%This paper presented a unified approach for monitoring heterogeneous Hw/Sw platforms based on the Performance Application Programming Interface (PAPI). The proposed approach offers to Sw developers the support to design and implement Coarse-Grain Virtual Reconfigurable Circuits instrumented with custom monitors and integrated into a Processor-Coprocessor System.  Specifically, following the \textsc{Papify} instrumentation syntax, both Sw and Hw monitoring information have been retrieved in a transparent way. Experimental evaluation demonstrates the effectiveness and the advantages of the proposed approach. Using this monitoring has an impact on the application performance of 6.20\%, due to the fact that information are written into csv files. This overhead would be potentially reduced when the monitoring information will be directly fed to the adaptation manager and not written into a file.

This paper presented the combination of \textsc{Papify} and MDC tools to support Sw developers in the design and implementation of Coarse-Grain Virtual Reconfigurable Circuits instrumented with custom monitors and integrated into a Processor-Coprocessor System. The users specify the applications to be accelerated as dataflow specifications, and the automated toolchain deploys the final system together with the  Application Programming Interfaces (APIs) to mask the processor-coprocessor communication. The combination of these APIs and of \textit{Sw Libraries} provided by \textsc{Papify} offer a unified approach for monitoring heterogeneous Hw/Sw platforms based on the Performance Application Programming Interface (PAPI), in which both Sw and Hw monitoring information can been retrieved in a transparent way.

Experimental evaluation demonstrates the effectiveness and the advantages of the proposed approach. Using this monitoring has an impact on the application performance of 6.20\%, due to the fact that information are written into csv files. Indeed, the proposed approach is one of the mandatory steps to implement self-adaptive systems, but at the moment the collected data are not automatically fed-back to any adaptation manager. When the monitoring information will be directly fed to the adaptation manager and not written into a file, this overhead would be potentially reduced and the system would be self-adaptive.

For instance, in the use-case adopted for the assessment, monitoring the latency would help to verify that the application is working properly. Once estimated, at the design time, that the application should complete in $x$ clock cycles, if after $3x$ the execution is not complete and the number of output tokens is less than expected, the manager could assume that the execution stacked for any reason and decide to reset and restart the application.

As a follow up, and to deal both with the overhead impact and the analysis of the information, all the monitoring data will be feeding SPIDER. By doing so, the low-level hardware information could be used to improve the workload distribution decision making carried out by this manager. This last step will close the loop and enable the self-adaption of a heterogeneous parallel system, integrating monitoring, decision making and reconfiguration capabilities~\cite{Palumbo_2019x2}.

%the system depending on information about the main key performance indicators (such as energy, latency and throughput). An on-going work is related to the models of architecture for latency, starting from the work, based on energy, presented in~\cite{Pelcat2017}. \textcolor{red}{FP: I would remove this closing sentence that it is more related to the run-time manager that is not the focus of this work and I would say that this work is an important step forward to allow closing the loop presented in SAMOS. You can put the citation as to appear in proceeding of SAMOS 2019. To me talking about the MoA here is far less relevant than the adaptation loop.}

\section*{Acknowledgments}
%\noindent \footnotesize \textbf{Acknowledgments}:
This research received funding from the European Union under grant agreement No 732105 (CERBERO H2020 project) and No 783162 (FitOptiVis ECSEL project), from the Sardinian Regional Government within the PROSSIMO project (POR FESR 2014/20-ASSE I) and from the Ministry of Economy and Competitiveness of the Spanish Government through the PLATINO project, No. TEC2017-86722-C4-2-R. The authors would like to thank the Universidad Polit\'ecnica de Madrid for its support under the Programa Propio RR01/2016 predoctoral contract. %Furthermore, this work has been funded by the Sardinian Regional Government within the PROSSIMO project (POR FESR 2014/20-ASSE I).

%%% Please leave the following line as is and insert your bibliography items below.

\end{document}